\documentclass[]{interact}

\usepackage{amsmath, amsthm, amssymb}
\usepackage[english]{babel}
\usepackage{dsfont}
\usepackage{mathdots}
\usepackage{graphicx}
\usepackage{enumerate}
\usepackage{color}
\usepackage{float}
\usepackage{placeins}
\usepackage{url}
\usepackage{comment}
\usepackage{todonotes}
\usepackage{rotating}
\usepackage{multirow}
\usepackage{diagbox}
\usepackage{ulem}
\usepackage{harvard}

\newtheorem{thm}{{Theorem}}
\newtheorem{lem}[thm]{{Lemma}}

\newcommand{\var}{\mbox{Var}}
\newcommand{\cv}{\mbox{CV}}

\newcommand{\RKI}{\mbox{RKI}}
\newcommand{\E}{\mbox{E}}

\newcommand{\N}{\mathds{N}}

\newcommand{\COVID}{COVID-19}
\newcommand{\SARS}{SARS-CoV-2}

\newcommand{\ind}{\mathds{1}}

\allowdisplaybreaks
\begin{document}
\title{Estimation of the Distribution of the Individual Reproduction Number: The Case of the {\COVID} Pandemic}

\author{
\name{A. Braumann\textsuperscript{a}, J. Krampe\textsuperscript{b}, J.-P. Kreiss\textsuperscript{a}  and E. Paparoditis\textsuperscript{c}}
\affil{\textsuperscript{a} TU Braunschweig, Germany; \textsuperscript{b} University of Mannheim, Germany;
\textsuperscript{c} University of Cyprus, Cyprus}
}
\maketitle

\begin{abstract}
    We  investigate  the problem of estimating the distribution of the individual reproduction number governing  the {\COVID} pandemic. Under the assumption  that this random variable follows a Negative Binomial distribution, we focus on constructing  estimators  of the parameters of this distribution using reported infection data and   taking into account issues like  under-reporting or the time behavior of the infection and of the reporting processes. To this end, we extract information from  regionally dissaggregated data reported by German 
    health authorities, in order to estimate not only the mean but also the variance of the distribution of the individual reproduction number. In contrast to the mean, the latter parameter also depends on the unknown under-reporting rate of the pandemic. The estimates obtained   allow not only for a better understanding of the time-varying behavior of the expected value of the individual  reproduction number but also  of its dispersion, for the construction of bootstrap confidence intervals  and  for a discussion  of  the  implications of different  policy interventions. Our methodological investigations are accompanied  by an empirical study of the development of the {\COVID} pandemic in Germany, which  shows a strong overdispersion of the individual reproduction number. 
\end{abstract}


\vspace*{0.5cm}

\section{Introduction}
\label{intro}    
The individual reproduction number $R$ is commonly used in epidemiology to quantify the 
transmission of a disease.
$R$ describes the number of secondary
infections caused by a single {\SARS}-positive individual. The random variable $R$ is a very 
important quantity in controlling the {\SARS}-pandemic. 
Of special interest is the expectation $\E (R)$ of the reproduction number, often denoted as $R_0$ 
and called {\sl basic reproduction number}. A value
$R_0<1$ indicates that the pandemic is under control, while $R_0>1$ 
indicates a strong warning
that the pandemic is in an exponential growth stadium. 
Notice that the expectation $\E (R)=R_0$ 
is only one parameter of the distribution of the random variable $R$. 
Even though we treat $R$ as a random variable, the reproduction number also plays an important 
role in deterministic modeling in epidemiology which typically is based on ordinary differential
equations including the Kermack-McKendrick epidemic model SIR (Susceptible-Infectious-Removed) 
and the SEIR (Susceptible-Exposed-Infectious-Removed) model. For an introduction to the reproduction
number $R$ and especially to the Basic Reproduction Number $R_0$ we refer to 
\citeasnoun{ChowellBrauer2009}. 

Estimates for $R_0$ or $R_{0,t}$, if possible changes over time are taken into account,  
on the basis of observed non in-depth case numbers can be found in many papers in the literature. 
A fundamental alternative would be to estimate $\E (R)=R_0$ from in-depth tracking of infection-chains.
\citeasnoun{Dehningetal2020b} discuss a model-free estimation of the reproduction number $R_{0,t}$ 
and compare it with the standard techniques applied by
the Robert Koch Institute (RKI), which is the German government's central scientific institution
in the field of biomedicine.
Quite important for the various RKI-estimators
is the so-called generation duration or generation time.
It should be noted, that it is most difficult to estimate the reproduction number $R_t$ or even its expectation $R_{0,t}$ properly, at change points of the 
transmission (spread) of the virus, triggered for example by specific 
countermeasures like social distancing or efficient cluster tracing with prompt
isolation.

We will follow the approach of developing  methodology on the basis of non in-depth reported
case numbers but we will not only focus on the expectation but also on the entire
distribution of the reproduction number $R$.
\citeasnoun{Corietal2013} and also \citeasnoun{LloydSmithetal2005} together with the associated supplementary material (\citeasnoun{LloydSmithetalSuppl2005})
suggested a Negative Binomial distribution to model the stochastic
behavior of $R$. For {\COVID}-pandemic data, the Negative Binomial distribution is also used in \citeasnoun{Althouse2020} and \citeasnoun{Endo2020}. Of special interest is the ability of the Negative Binomial distribution
to include possible dispersion, which is
rather likely to be present in the {\COVID}-pandemic.
Dispersion means that the standard deviation or variance of a distribution
may vary independently of the mean. The latter for example is not possible for the also often used
Poisson distribution, for which variance and mean coincide. 
We refer to \citeasnoun{Azmonetal2014} for 
Poisson modeling when describing methodology to estimate the reproduction
number $R$. It is worth mentioning that \citeasnoun{Azmonetal2014}  
also consider estimates of $R$ if under-reporting is present.

Strongly related to dispersion, which
we will define more precisely in the next section, is the several times
by epidemiological experts expressed fear
that only (very) few super-spreading events with a very high number of secondary infections could drive 
the {\COVID}-pandemic into a critical state. This may even be the case, when some 80 per cent of 
newly infected individuals in fact lead to none or only one secondary infection and even 
when the expectation $R_0$ is close to one.

In this paper we discuss and investigate the opportunities to estimate the distribution of the 
individual reproduction number $R$ for the {\COVID}-pandemic in Germany on the
basis of non in-depth infection data provided on a daily basis by the Robert Koch Institute (RKI).
Cumulative data about newly reported cases, totally infected cases, fatalities as well as
a 7-days incidence rate per 100,000 inhabitants can be found on the website 
\url{http://corona.rki.de}.
The reported cases are based on positive laboratory testing of {\SARS} and are 
denoted as {\COVID}-cases irrespective of {\COVID}-symptoms.
The data is available separately per local states 
(Bundesl\"ander), districts (Landkreise), age-groups and gender, to name only a few.
\begin{figure}[htbp]
\begin{center}
 \includegraphics[scale=0.5]{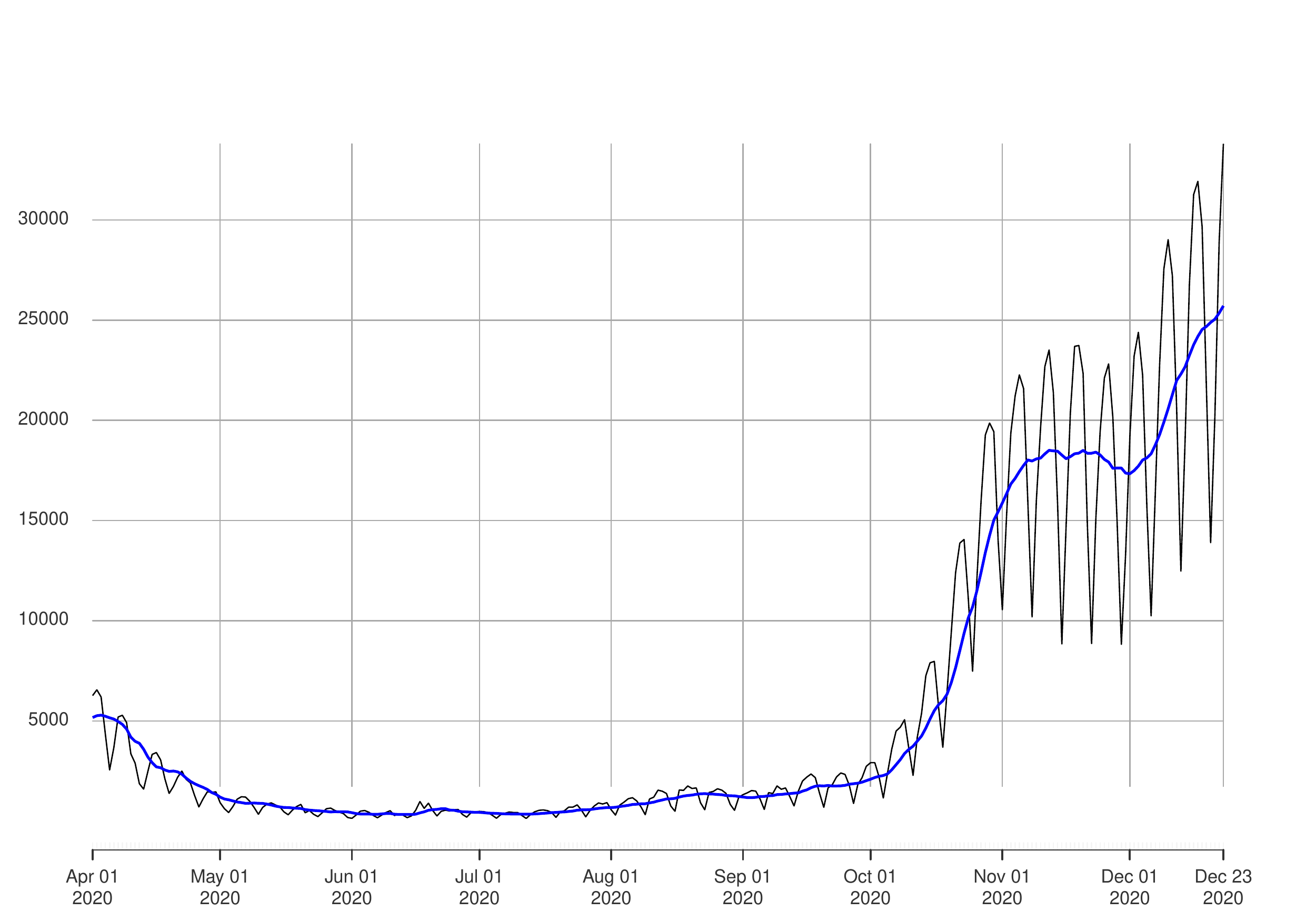}
\caption{Daily numbers of reported \COVID-cases, denoted by $\RKI_t$, in black. A moving average over seven days of $\RKI_t$ is given in blue.}
\label{fig-01}
\end{center}
\end{figure}
When considering daily {\COVID}-cases it is important to carefully distinguish 
for individuals
the time $t$ at which an {\SARS}-infection took place, the time 
at which the
case was first reported to the (local) health authorities (if it was 
laboratory confirmed at all)
and the time at which the case finally was reported to RKI. \\
For the investigations in this paper we typically consider times 
$t$ at which an
infection with {\SARS} takes place. 
It is most likely that the difference of the time of infection 
of a case, which in the end will be reported, and the
time of reporting follows a distribution over a couple of days. 
On average, we assume a time shift of $\tau=7$ days, which seems to be reasonable.
$\tau$ consists of the number of days 
(approximately $4$) between a {\SARS} infection took place and the 
onset of symptoms plus 
the time that elapses (after onset of symptoms) until a PCR-test is carried out 
and a (positive) result becomes known (approximately another $2$ days) plus the time 
(approximately another day) that elapses until the case is first reported to 
the health authorities. It is important to know
that in Germany it is mandatory that for a person
for whom a proof of the pathogen \SARS \, is obtained (mostly by a PCR-test),
a report has to be transmitted to the local health department
within 24 hours.
After this first reporting time to the health authorities the reporting chain  
is continued via the responsible state health authorities to the Robert 
Koch Institute, which will take further days. \\
For this paper the relevant times 
are the infection time and the time of the first reporting.
We denote by $\RKI _s$ the number of \COVID -cases first 
reported at time $s$ to the health authorities. Then, roughly speaking
the number of infections at time $t$, which in the end will be reported, and
$\RKI _{t+\tau}$ are related to each other. \\
Figure \ref{fig-01} displays the daily data $\RKI _t$ for the second, third and fourth quarter of 2020
together with a moving average smoothing over $7$ days. Averaging over $7$
days seems to be appropriate since a very strong periodic behavior
of $\RKI _t$ over the week is observed. The reason for this most likely lies
in the fact that the reporting chain between weekdays and weekends is
substantially different.
We argued that it is important to estimate not only the expectation 
but also the dispersion or equivalently the variance of the individual 
reproduction number $R$, for example in order to assess 
whether or not this quantity changes over time. 
To do so we present a setup of assumptions under which we are able 
to give reasonable estimates
of the variance (or dispersion) of $R$. 
To this end it appears to be necessary to rely on {\COVID}-data on district
(Landkreis) level in Germany. 

As a result it will be seen that the parameters of the distribution of $R$ 
indeed have changed over time and that
dispersion (even over-dispersion) is rather relevant. We will make the 
implication of this clear
and we also report upon simulation results which describe the changes 
in the behavior
of the development of the pandemic if the rather relevant super-spreading events are (to some extend) 
quickly discovered by the health authorities with the result that 
a substantial part of the secondary infected individuals in such a 
super-spreading event can promptly be isolated.\\
There exist already numerous papers in the literature which investigate 
the effects of social distancing or contact bans and also
of cluster tracing within a variety of models. 
\citeasnoun{Dehningetal2020a} used MCMC-sampling
in a SIR-model and obtained that already in case of mild social distancing 
a substantial reduction of the spreading rate is achieved.
\citeasnoun{Contrerasetal2020} describe that contact tracing is very effective 
in stabilizing the number of infections (i.e. the
observed number of newly infected individuals approaches a constant value), 
while inefficient contact tracing leads to an
approximately exponential growth of the number of infections. Moreover, 
the paper emphasizes that a combination of symptom-based
testing together with effective contact tracing appears to be most effective.
Additionally 
\citeasnoun{Lindenetal2020} describe clearly
that a break-down of a TTI-strategy (Test-Trace-Isolate) will lead to an 
increasing number of not-reported {\COVID}-cases
and therefore accelerates the spread of the virus.

The paper is organized as follows. In the next section we describe our 
modeling and state the main
assumptions used for our results. Section 3 presents the proposed estimators 
of the mean and of the variance of the individual reproduction number  taking 
into account under-reporting and the time behavior of the infection and of 
the reporting chain. Section 4 introduces Bootstrap-based confidence intervals for $R_0$. Section 5 examines the fit of the distributional 
assumptions made, presents estimators describing the development of 
the {\COVID} pandemic in Germany, and studies effectiveness of cluster tracing.  


\section{Modeling and Assumptions}
\label{model}

As already mentioned,  the Negative Binomial distribution has been suggested in the literature as an
appropriate  model to describe the randomness of the reproduction number $R$. The Negative Binomial distribution with parameters $p\in (0,1)$ and
$r \in \mathbb N$ is well-known in statistics for modeling the number of failures in a sequence of independent  Bernoulli($p$)-trials
until the $r$-th success occurs. 
It has been extended to parameters
$p\in (0,1)$ and $r \in (0,\infty )$ (we use the abbreviation $\mathcal{NB}(p,r)$) via a consideration of a classical
Poisson-distribution with random parameter $\lambda $ distributed according to a specific 
Gamma-distribution. To be precise a random variable $R$ possesses a (generalized) 
Negative Binomial distribution $\mathcal{NB}(p,r)$, if and only if 
\begin{equation}
\label{eq-01}
P_{p,r}(R=k)= \int _0^{\infty} e^{-\lambda } \frac{\lambda ^k}{k!}\, f_{r,\frac{p}{1-p}}(\lambda )\, d\lambda,\ \ k\in \mathbb N_0,
\end {equation}
with 
\[  f_{r,\frac{p}{1-p}}(\lambda )= \frac{1}
{\Gamma (r)}\big(\frac{p}{1-p}\big)^r \lambda ^{r-1}
e^{-\lambda \frac{p}{1-p}},
\]  
the density of a Gamma-Distribution with parameters $r\in (0,\infty)$, 
$p/(1-p) \in (0,\infty )$ and $ \Gamma(r)=\int_0^1x^{r-1}e^{-x}dx$ the Gamma function.

The Negative Binomial distribution as a generalized Poisson
distribution allows for a more flexible modeling of rare events with different expectation and variance.
For $R\sim \mathcal{NB}(p,r)$ we have
\begin{equation}
\label{eq-02}
\E (R)= \frac{r\, (1-p)}{p} \ \ \text{\rm and } \ \ \var(R)= \frac{r\, (1-p)}{p^2}.
\end {equation} 
Typically the expectation $\E (R)$ is denoted by $R_0$. 
The Negative Binomial distribution possesses a coefficient of variation 
$\cv (R)=\var (R)/\E (R)=1/p$, and allows for modeling the distribution of the reproduction number 
$R$ with dispersion. \\
The dispersion parameter $\kappa$ is defined through
\begin{equation}
\label{eq-03}
\var (R) = R_0\cdot \Big(1+\frac{R_0}{\kappa}\Big),
\end {equation} 
which leads to $\kappa =r$ (cf. \citeasnoun{LloydSmithetal2005}). 
The dispersion parameter $\kappa $ and  the coefficient of variation
both are widely used to quantify the size of the variance given the expectation
of a random variable. For fixed $\E (R)=R_0$ we have that the
smaller $r$ the larger is the variance of the reproduction number $R$.
To see that $\kappa $ and $\cv$ behave rather the same 
let us assume two sets of parameters $(p,r)$ and $(\widetilde p, \widetilde r)$
of Negative Binomial distributions describing the random behavior of two
reproduction numbers $R$ and $\widetilde R$ 
(with dispersion parameters $\kappa$ and $\widetilde \kappa$). 
If we fix $\E R=\E \widetilde R=R_0$ then
\begin{equation}
\label{eq-03a}
\frac{\cv(\widetilde R)}{\cv (R)} = \frac{\kappa \cdot(R_0+\widetilde \kappa)}{\widetilde \kappa \cdot (R_0+\kappa )},
\end {equation} 
which 
is a decreasing function in $\widetilde \kappa$.

For Negative Binomial distributions it is known that 
$r\to \infty $ and $p\to 1$, such that $r(1-p) \to \lambda \in (0,\infty)$, 
leads to a classical Poisson($\lambda$)-limit, 
while for $r=1$ the  Negative Binomial distribution
coincides with the geometric distribution with parameter $p$.

We will model the number of secondary infections caused by an individual {\COVID}-case
with infection time (day) $t$, i.e., the reproduction number $R_t$ at time $t$, 
via a $\mathcal{NB}(p_t,r_t$)-distribution. 
If we further denote the number of 
newly infected cases at time $t$ by $N_t$, we then are faced with a total of
\begin{equation}
\label{eq-04}
\sum _{i=1}^{N_t} S_{i,t},
\end {equation} 
secondary infections in the future. Here $S_{1,t}, S_{2,t}, \ldots $ denote i.i.d. random variables 
distributed according to $\mathcal{NB}(p_t,r_t$). It should be noted that the random variables $S_{i,t}, i=1, 2, \ldots ,$ are i.i.d. copies of
$R_t$. In order to avoid confusion we decided to use the notation
$R_t$ for the generic 
reproduction number, while we label
the random secondary infections from a single infected individual by $S_{i,t}$.

For several reasons not all of these future cases will be reported to the German health authorities and
subsequently will not show up in the RKI-statistics of newly laboratory-confirmed {\COVID}-cases.
One major, but not the only reason for this under-reporting is 
that a substantial number of
{\SARS}-infections are asymptomatic.
The studies \citeasnoun{Buitrago2020} and 
\citeasnoun{OranTopol2020} report that percentages of 
asymptomatic cases vary between 20\% to 45\%. 
These values coincide with results from a study in the 
community of Kupferzell (Germany), where a percentage of
asymptomatic cases of 24.5\% has been observed, cf.
\citeasnoun{Santosetal2020}.\\
Although it seems difficult to assess the exact rate of
under-reporting, this rate is of course a relevant quantity when
investigating the development of the pandemic.
Some studies state rates of not reported cases up to 80\%. 
See for example \citeasnoun{Streecketal2020}, where a 5-fold
higher number of infections than the number of
officially reported cases for a specific community in Germany 
with a super-spreading event is reported or the already cited
study in Kupferzell (\citeasnoun{Santosetal2020}), where is was
observed that six times more adults than reported have 
been infected, which is a rate of under-reporting of 
80\% or even higher. This coincides with statements made by 
the  RKI, that the number of
infected individuals approximately is 4-6 times higher than
the number of reported cases. Besides this, the RKI stated that
there is no evidence for a substantial change of this factor
over the last months. Finally, \citeasnoun{Rahmandadetal2020},
in a study across 86 nations, found out that
under-reporting varies substantially over countries. For Germany,
\citeasnoun{Rahmandadetal2020} estimated a ratio of 
actual to reported cases of about 6 to 7. \\
In this paper we denote the proportion of {\COVID}-cases reported
to the German health authorities by
$p_{0,t}$, and allow this rate to vary (slowly) with time. A value
of $p_{0,t}\approx 0.2$ seems realistic in the light of the
studies cited.\\
From a number of $N_t$ newly infected individuals at time $t$ 
we therefore will see within the 
statistics of RKI only a binomial thinned selection of $N_t$, which we denote by $N_t^{\prime }$.
According to our assumption of a reporting rate of $p_{0,t}$ and because of a not small number of cases
it is reasonable to assume that, approximately, 
\begin{equation}
\label{eq-04aa}
\frac{N_t^{\prime}}{N_t} \approx p_{0,t}.
\end {equation} 
It is worth mentioning
that the number of reported cases $N_t^{\prime}$ out of 
infections happened at time $t$ do not show
up within the RKI-statistics neither at time $t$ nor at a 
single time point in the future. Rather, 
occurrence in the RKI-statistics will spread over a span 
 of days.

This further means that from the total number
$\sum _{i=1}^{N_t} S_{i,t}$  of secondary infections caused 
by $N_t$ primary infections we only observe
$\sum _{i=1}^{N_t} \widetilde S_{i,t}$ laboratory-confirmed 
cases with the statistics of RKI, where 
$\widetilde S_{i,t}$ possesses a Binomial-distribution with
parameters $S_{i,t}$ and $p_{0,t}$.
Equivalently, the number of reported {\SARS} secondary 
infections out of a cohort of $N_t$ primary
infected individuals can be written as
\begin{equation}
\label{eq-05}
\sum _{i=1}^{N_t} \sum _{j=1}^{S_{i,t}} Z_{i,j},
\end {equation} 
where $(Z_{i,j}, i,j \in \mathbb N)$ is a family of i.i.d. Bernoulli($p_{0,t}$)-distributed random variables.
Success, i.e. $Z_{i,j}=1$, means that a secondary infected individual gets a positive {\COVID}-test 
at some day in the future. As stated above
it is of course not realistic to assume that all 
$\sum _{i=1}^{N_t} \sum _{j=1}^{S_{i,t}} Z_{i,j}$ cases will show up in the RKI-statistics at one single
day. Rather occurrence of these cases in the RKI-statistics will 
spread over a span of some days.
 
Before further elaborating on  this point let us take a look at the distribution of the total numbers of secondary infections 
$\sum _{i=1}^{N_t} S_{i,t}$ and reported secondary infections
$\sum _{i=1}^{N_t} \widetilde S_{i,t}$. Since it is known that the
Negative Binomial distribution is additive, we immediately obtain, assuming independence of the 
single cases, that
$\sum _{i=1}^{N_t} S_{i,t}\sim \mathcal{NB}(p_t, N_t\cdot r_t)$. 
Moreover, $\widetilde S_{i,t}$ is a Binomial-thinning of $ S_{i,t}$. The property that Binomial-thinning
changes the parameter but not the family of Poisson-distributions carries over to the family of
Negative Binomial distributions, cf. Lemma 1 of the Appendix. This means that we end up with
\begin{equation}
\label{eq-06}
\widetilde S_{i,t} \sim \mathcal{NB}(q_t, r_t), \ \ i=1,2, \ldots ,
\end {equation} 
with 
\begin{equation}
\label{eq-06aa}
q_t= \frac{p_t}{p_{0,t}+p_t-p_{0,t}\cdot p_t} .
\end {equation} 
The parameter $q_t$ depends on the hardly 
known percentage $p_{0,t}$ of {\SARS}-infections reported to the health authorities.
Furthermore, 
\begin{equation}
\label{eq-06a}
\sum _{i=1}^{N_t} \widetilde S_{i,t} \sim \mathcal{NB}(q_t, N_t\cdot r_t).
\end {equation}
So far we focused on time points $t$ at which {\SARS}-infections take place.
We already discussed in the introduction that these
time points $t$ should not be confused with the time points at which 
\SARS -infections are first reported to the health authorities 
(recall that we denoted the number of {\COVID}-cases first reported at time point $t$
to the health authorities by $\rm{RKI}_t$). 
We argued that there is a (random) time shift between these two time points with a
likely mean of $\tau = 7 $.

So as not to make things too complicated and still take time shifts as well as
random fluctuations of reporting delays over a span of days into account,
we make the following two assumptions
\begin{equation}
\label{eq-07}
\sum _{s=0}^{6} N_{t-s}^{\prime} \approx \sum _{s=0}^{6} \rm{RKI}_{t+\tau -s}.
\end {equation} 
and
\begin{equation}
\label{eq-07-2}
\sum _{s=0}^{6} N_{t-s}^{\prime} \approx \sum _{s=0}^{6} \sum _{i=1}^{N_{t-4-s}} 
\widetilde S_{i,t-4-s}.
\end {equation} 
The first assumption means that newly infected cases, which are of a type 
that will be
reported in the future, summed up over a week approximately will occur in the RKI-statistics also within a week but shifted by $\tau$ days to the future, 
while the second assumption is a relaxation of 
$N_t^{\prime } \approx \sum _{i=1}^{N_{t-4}} \widetilde S_{i,t-4} $. 
The latter assumption would mean that secondary infections occur with a 
fixed time delay of $4$ days to the primary infection. Instead of such a strict
assumption \eqref{eq-07-2} means that the two quantities are roughly the same
if they are summed up over a week.
Here the number $4$ can be viewed as generation time of the virus.\\

\section{Estimation of Parameters of the Distribution of the Reproduction Number}
\label{Estimation}

Based on the assumptions made in the previous section
it follows that the smoothed estimate  of the mean of the  reproduction number
published on a daily basis by RKI, and denoted by $\widehat R_{0,t}^7$ fulfills
\begin{equation}
\label{eq-08}
\widehat R_{0,t}^7:=
\frac{\sum _{s=0}^{6} \rm{RKI}_{t-s}}{\sum _{s=0}^{6} \rm{RKI}_{t-4-s}}
\approx
\frac{\sum _{s=0}^{6} N_{i,t-\tau-s}^{\prime} }{\sum _{s=0}^{6} N_{t-4-\tau -s}^{\prime}}
\approx
\frac{\sum _{s=0}^{6} \sum_{i=1}^{N_{t-4-\tau-s}}\widetilde S_{i,t-4-\tau-s} }{\sum _{s=0}^{6} N_{t-4-\tau -s}^{\prime}},
\end {equation} 
cf. \eqref{eq-07} and \eqref{eq-07-2}.
It is often noted that $\widehat R_{0,t}^7$ reflects the reproduction behavior
approximately $14$ days ago. 
To understand this, notice that we need two generations of \SARS -infected
individuals in order to be able to compute reproduction numbers.
Since the generation time of \SARS{}  is approximately $4$ days 
and since the proposed computation of the reproduction number
uses a left-sided smoothing over $7$ days (cf. \eqref{eq-08}), 
which leads to a time-shift of $3$, a reasonable reproduction number 
can only be calculated with a delay of $4+3$ days after primary 
infections have taken place.
Because $\widehat R_{0,t}^7$ in \eqref{eq-08} is computed on the basis of 
reported case numbers we have to face on top the aforementioned 
general reporting delay of $\tau$, for which $\tau=7$ seems realistic. 
In total for $\widehat R_{0,t}^7$ we end up with a delay of about $14$ days
in describing the reproduction behavior of \SARS .

From \eqref{eq-06a} we have that the distribution of the numerator
$\sum _{s=0}^{6} \sum_{i=1}^{N_{t-4-\tau-s}}\widetilde S_{i,t-4-\tau-s} $, given the numbers $N^{\prime}_{t-4-\tau -s}, s=0,\ldots , 6,$ 
approximately is (assume that $q_t$ and $r_t$ only vary slowly over time and therefore 
$q_{t-4-\tau-s}\approx q_{t-\tau-7}$ and $r_{t-4-\tau -s}\approx r_{t-\tau-7}, s=0,\ldots ,6$)
\begin{equation}
\label{eq-09}
\mathcal{NB}(q_{t-\tau-7} , r_{t-\tau -7}\cdot \sum _{s=0}^6 N_{t-\tau-4-s} ),
\end {equation} 
with (conditional on $ \sum _{s=0}^6 N_{t-\tau-4-s} $) expectation
\begin{equation}
\label{eq-10}
\sum _{s=0}^6 N_{t-\tau -4-s} \cdot \frac{r_{t-\tau-7}\cdot (1-q_{t-\tau-7})}{q_{t-\tau-7}}.
\end {equation} 
Using the further approximation from \eqref{eq-04aa} this leads to the following value 
that is estimated by $\widehat R_{0,t}^7$ 
\begin{equation}
\label{eq-11}
\frac{\sum _{s=0}^6 N_{t-4-\tau-s}}{\sum _{s=0}^6 N_{t-4-\tau-s}^{\prime}}
\cdot \frac{r_{t-\tau-7}\, (1-q_{t-\tau-7})}{q_{t-\tau-7}}
=\frac{1}{p_{0,t-\tau-7}}\cdot \frac{r_{t-\tau-7}\, (1-q_{t-\tau-7})}{q_{t-\tau-7}}.
\end {equation} 
Fortunately, we obtain by  simple algebra and using \eqref{eq-06aa}, that
\begin{equation}
\label{eq-13}
\frac{1}{p_{0,t-\tau-7}}\cdot \frac{r_{t-\tau-7}\, (1-q_{t-\tau-7})}{q_{t-\tau-7}}
=
\frac{r_{t-\tau-7}\, (1-p_{t-\tau-7})}{p_{t-\tau-7}},
\end {equation} 
which is the expectation $R_{0,t-\tau-7}$ of the reproduction number $R_{t-\tau-7}$ we are
interested in.

In order to be able to estimate both parameters $r$ and $p$ of a Negative Binomial fit to the 
distribution of the reproduction number $R$ we further need an estimator of 
$\var (R)$. For this we
need somehow replicates of realizations of $R$, which we will obtain from reported {\COVID}-cases 
on district level from Germany. In total, Germany is divided into 
about $401$ districts with
population numbers ranging from $34,193$ to $3,669,491$. For each district RKI provides daily {\COVID}-cases along the same
guidelines as for Germany as a whole. As before we denote the number of newly 
infected (not necessarily reported!)
{\COVID}-cases by $N_{t,\ell }$, where $t$ counts the day (time) and 
$\ell =1, \ldots, L=401$ denotes the number
of the district. For a single primary infected individual we assume that the total number of secondary infected
persons possesses a $\mathcal{NB}(p_t, r_t)$-distribution, which does not depend on the district.   The total
number of secondary infections caused by $N_{t,\ell }$ primary infected individuals then follows a
$\mathcal{NB}(p_t, N_{t,\ell}\cdot r_t)$-distribution. 
Following the arguments in Section \ref{model} we obtain that the total number of (in the end) reported {\SARS}-
secondary infections out of a number of $N_{t,\ell}$ primary infections in district $\ell$, which
we denote by $N_{t,\ell}^{\prime}$,
is distributed according to $\mathcal{NB}(q_t, N_{t,\ell}\cdot r_t)$, cf. \eqref{eq-09} and \eqref{eq-08}. Note that if there is evidence that the infection rates of some districts differs highly from the infection rates of the others then a subset selection might be preferable.

In order to relate the number of daily first reported cases 
$\rm{RKI}_{t,\ell}$ within district $\ell$ with
the total number of secondary infections  $N_{t,\ell}^{\prime}$, for which the reporting is spread over
some of days, we assume in accordance with \eqref{eq-07} and  \eqref{eq-07-2} for each district 
$\ell =1, 2, \ldots, L$
\begin{equation}
\label{eq-14}
\sum _{s=0}^{6} N_{t-s,\ell}^{\prime} \approx \sum _{s=0}^{6} \rm{RKI}_{t+\tau -s,\ell},
\end {equation} 
and
\begin{equation}
\label{eq-15}
\sum _{s=0}^{6} N_{t-s,\ell }^{\prime} \approx \sum _{s=0}^{6} \sum _{i=1}^{N_{t-4-s,\ell }} 
\widetilde S_{i,t-4-s}.
\end {equation} 
Along the same lines as above (cf.  \eqref{eq-08}) the standard 7-days reproduction number of RKI,
i.e., for each $\ell =1, 2, \ldots , L$
\begin{equation}
\label{eq-16}
\widehat R_{0,t,\ell }^7:=
\frac{\sum _{s=0}^{6} \rm{RKI}_{t-s,\ell }}{\sum _{s=0}^{6} \rm{RKI}_{t-4-s,\ell }}
\end {equation} 
provides an estimator of $\E (R) =r_{t-\tau -7}\cdot \big(1-p_{t-\tau -7}\big)/p_{t-\tau -7}$, which
by our assumptions does not depend on $\ell$.

\begin{figure}[htbp]
\begin{center}
 \includegraphics[scale=0.5]{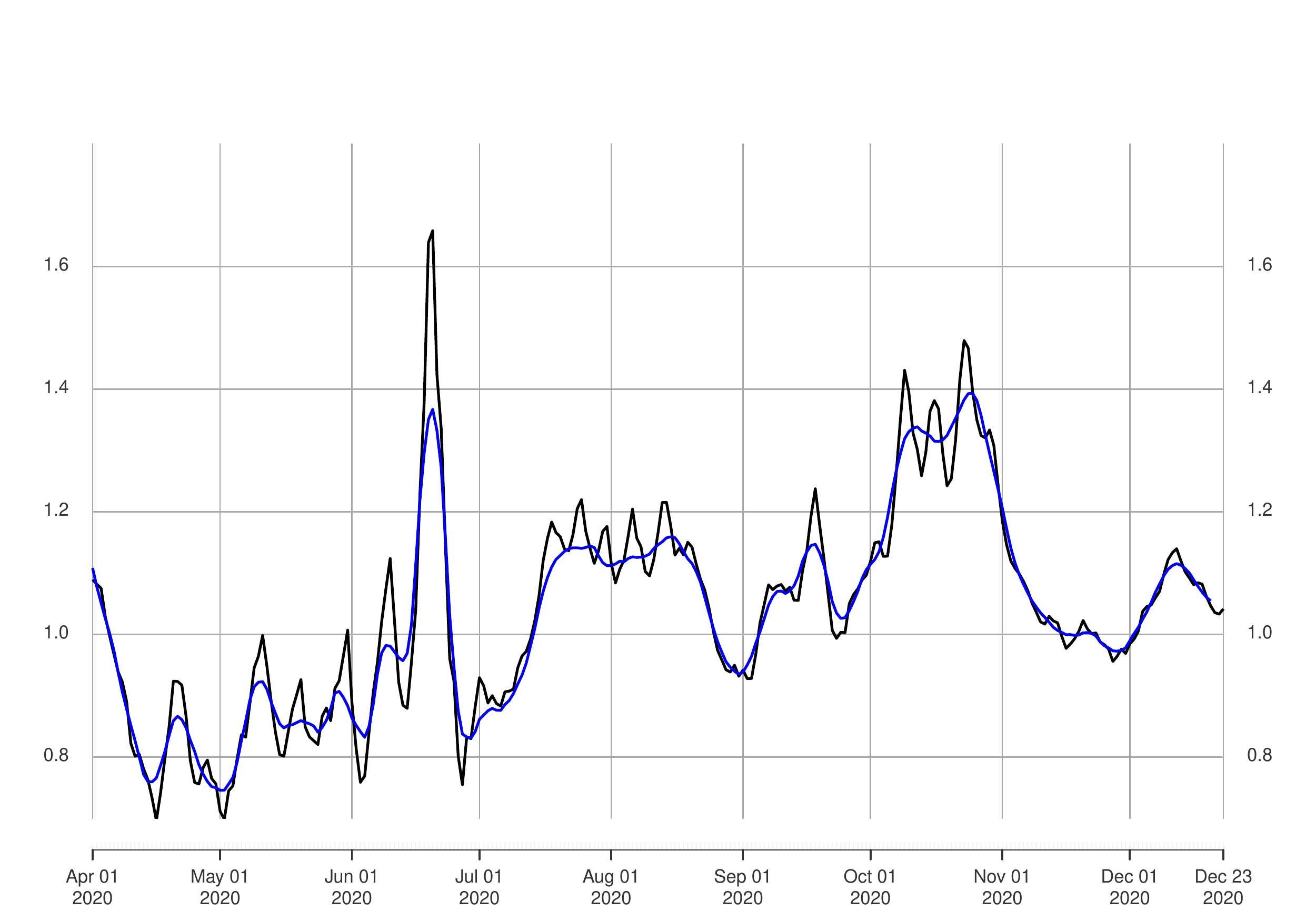}
\caption{The black line shows the 7-days smoothed RKI-estimator $\widehat R_{0,t}^7$. A moving average of $\widehat R_{0,t}^7$ of order $7$ is given in blue.}
\label{fig:R0hat}
\end{center}
\end{figure}

Averaging estimator \eqref{eq-16} over the districts and taking the heterogeneous variance into account, see \eqref{eq-variance-R0}, exactly leads to the 7-days smoothed RKI-estimator $\widehat R_{0,t}^7$ of the reproduction number based on reported {\COVID}-cases all over Germany (see Figure~ \ref{fig:R0hat}).

Our main focus, when turning to reported {\COVID}-cases on district level, is to obtain an estimator of the variance
$\var (R_{t-\tau-7}) = r_{t-\tau -7}\cdot \big(1-p_{t-\tau -7}\big)/p_{t-\tau -7}^2$. Because of the approximation of the
distribution of the numerator of  $\widehat R_{0,t,\ell }^7$ by a 
$\mathcal{NB}(q_{t-\tau-7} , r_{t-\tau -7}\cdot \sum _{s=0}^6 N_{t-\tau-4-s,\ell} )$-distribution
together with the approximation
\begin{equation}
\label{eq-17}
\frac{\sum _{s=}^6 N_{t-\tau -4-s}^{\prime}}{\sum _{s=0}^6 N_{t-\tau-4-s}} \approx p_{0,t-\tau-7},
\end{equation} 
cf. \eqref{eq-04aa} and also \eqref{eq-11},
we obtain for given $\sum _{s=0}^6 N_{t-\tau-4-s}=\sum_{s=0}^6 \rm{RKI}_{t-4-s,\ell}/p_{0,t-\tau-7}$, i.e., the denominator is considered fix,
\begin{align}
 \var(\widehat R_{0,t,\ell }^7)\approx 
 \frac{1}{\sum_{s=0}^6 \rm{RKI}_{t-4-s,\ell} \, p_{0,t-\tau-7}} r_{t-\tau -7} (1-q_{t-\tau-7})/q_{t-\tau-7}^2. \label{eq-variance-R0}
\end{align}
 This  means, the variance is heterogeneous among the districts with a factor $1/\sum_{s=0}^6 \rm{RKI}_{t-4-s,\ell}$. Taking this into account leads to the estimator \eqref{eq-22}, which is an estimator for $\var (\widetilde{S}_{i,t-\tau -7})/p_{0,t-\tau-7}$, i.e., the variance of the  number of {\sl reported} {\SARS}-secondary infection
cases from a single infected individual scaled by $1/p_{0,t-\tau-7}$.

\begin{equation}
\widehat{ \var (\widetilde S_{t-\tau-7})}:=
    \frac{1}{L} \sum_{\ell=1}^L \, \sum_{s=0}^6 \rm{RKI}_{t-4-s,\ell} \left(
    \widehat{R}_{0,t,\ell}^7 -  \widehat R_{0,t}^7\right)^2. \label{eq-22}
\end{equation}
Note that not necessarily an independence assumption across districts is required 
in order  for \eqref{eq-22} to be  a consistent estimator. Only consistency of  sample moments is required, which, for instance, can be achieved under some rather   weak dependence assumptions. This means that  neighboring districts may  be dependent but districts which are far apart from each other should behave nearly independent. 

\begin{figure}[htbp]
\begin{center}
 \includegraphics[scale=0.5]{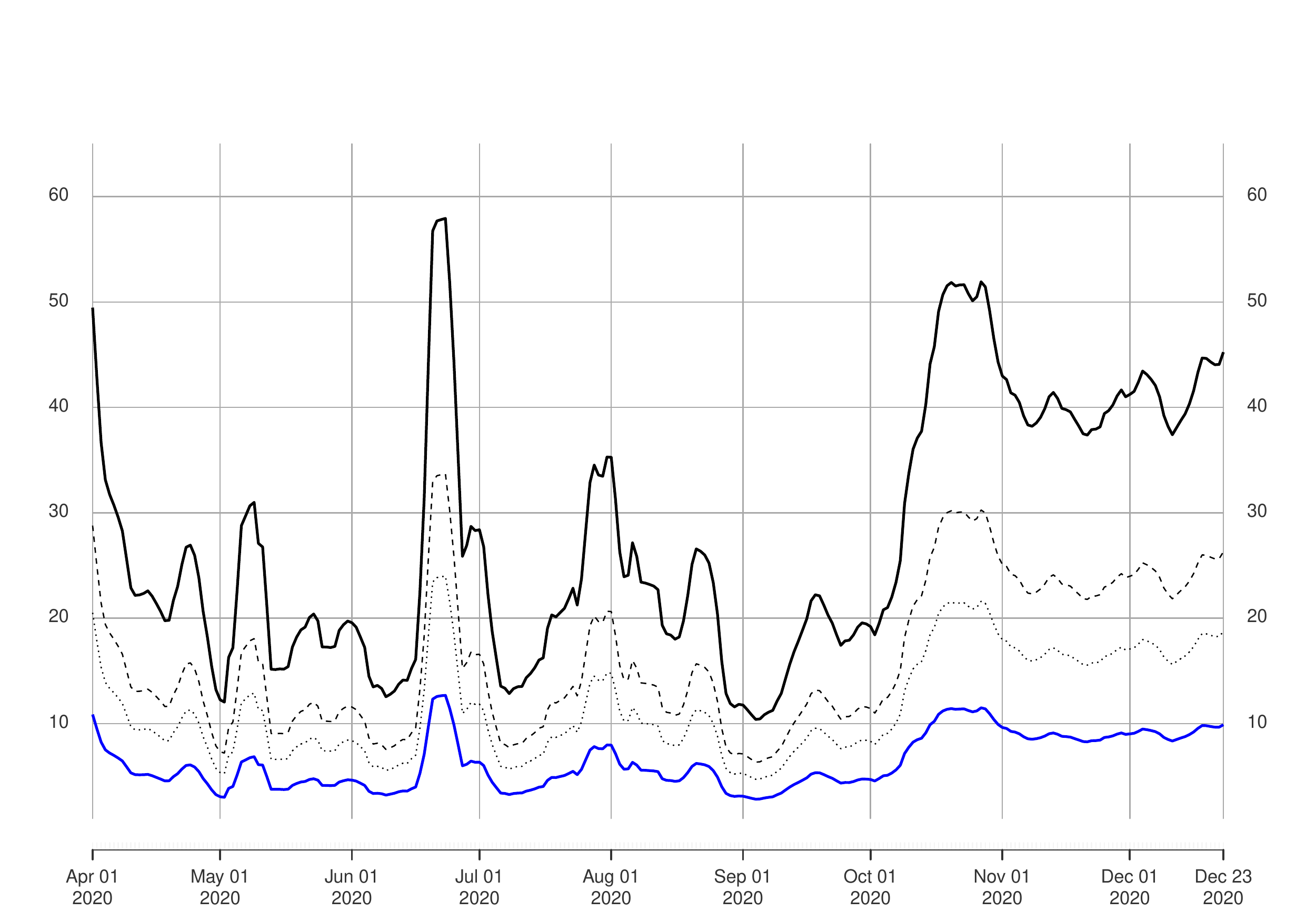}
\caption{The estimates of the variances $\var(\widetilde S_{t-\tau-7})/p_{0,t-\tau-7}$ (in blue) and $\var(R_{t-\tau-7})$ for $p_0=0.2$ (black solid line), $p_0=0.35$ (black dashed line), $p_0=0.5$ (black dotted line).}
\label{fig:variances}
\end{center}
\end{figure}

Since the distribution of  $\widetilde{S}_{i,t}$
is $\mathcal{NB}(q_t,r_t)$ we obtain 
\begin{align}
\label{eq-18}
\frac{1}{p_{0,t}}\cdot \var( \widetilde{S}_{i,t}) & = \frac{r_t \cdot (1-q_t)}{p_{0,t} q_t^2} \nonumber \\
& = \frac{r_t \cdot (1-p_t)}{p_t^2}\cdot (p_{0,t}+p_t-p_{0,t}p_t)\nonumber \\
&  = \var (R_t) \cdot (p_{0,t}+p_t-p_{0,t}p_t), 
\end {align} 
that is, for  $p_t>0$,
\begin{equation}
\label{eq-18b}
\var( R_t) = \frac{1}{(p_{0,t}+p_t-p_{0,t}p_t)}\,\cdot \frac{1}{p_{0,t}}
\var(\widetilde{S}_{i,t}).
\end {equation}
As it is seen, and in contrast to the expectation (cf. \eqref{eq-13}), the variance of the reproduction
number based on {\COVID}-cases reported to the health authorities
depends on the unknown reporting rate $p_{0,t}$. Since the reporting rate $p_{0,t}$ cannot be estimated
from reported data we only can calculate variance estimators for a variety of assumed 
reporting rates $p_{0,t}$ (see Figure \ref{fig:variances}).
Based on estimators $\widehat{\E R_{t-14}}:= \widehat R_{0,t}^7$, cf. \eqref{eq-08},
with $\tau =7$,
$\widehat{ \var (\widetilde S_{t-14})}$ as given in \eqref{eq-22} and because of
\eqref{eq-18b} together with explicit expressions of expectation and
variance of the Negative Binomial distribution $\mathcal{NB}(p_{t-14},r_{t-14})$
assumed for $R_{t-14}$, we finally are led to the following estimators
$\widehat p_{t-14}$ and $\widehat r_{t-14}$ for any fixed value of $p_{0,t-14}$
and the choice of $\tau =7$.
\begin{equation}
\label{eq-18c}
\widehat p_{t-14} =\frac{\widehat{R}_{0,t}^7 \cdot p_{0,t-14}}
     {\widehat{ \var (\widetilde S_{t-14})}-\widehat{R}_{0,t}^7 
     \cdot (1-p_{0,t-14})}
\ \ \text{and} \ \ 
\widehat r_{t-14} = \frac {\widehat{R}_{0,t}^7 \cdot \widehat p_{t-14}}
                          {1-\widehat p_{t-14}}\, .
\end {equation}

\section{Bootstrap Confidence Intervals}

Besides estimating the  mean  and the variance of  
the individual reproduction number  $ R_{t}$ as well as the parameters of the corresponding Negative Binomial distribution, it is important to construct confidence intervals for the unknown mean $R_{0,t}$. According to our previous discussion, the numerator of the estimator $ \widehat{R}_{0,t}^7$  approximately satisfies for $\tau = 7$
\begin{equation} \label{eq.boot1}
\sum_{s=0}^6 RKI_{t-s} \sim \mathcal{NB}\big(q_{t-14}, r_{t-14}\cdot \sum_{s=0}^6 RKI_{t-s-4}/p_{0,t-14}\big);
\end{equation}
see also the discussion before and after equation \eqref{eq-17} for the same property for observations obtained at the district level. Recall  that this distribution depends on the unknown under-reporting rate $ p_{0,t-14}$. Based on expression \eqref{eq.boot1} the following parametric bootstrap procedure can  be proposed for  constructing  a confidence interval for 
the mean $ R_{0,t-14}$ of the individual reproduction number.

\vspace*{0.3cm}
\noindent{\bf Bootstrap Algorithm, Confidence Intervals for $R_{0,t-14}$:}
\begin{enumerate}
\item[]\ {\it Step 1:} \  For    $p_{0,t-14}$  given and estimates $\widehat{q}_{t-14} $ and $\widehat{r}_{t-14}$, we generate for $t=15, 16, \ldots, n$,  pseudo random variables 
 $ (\sum_{s=0}^6 RKI_{t-s})^\ast $ 
 distributed as  
    \begin{equation} \label{bootS-1}  \big(\sum_{s=0}^6RKI_{t-s}\big)^\ast \sim \mathcal{NB}\big(\widehat{q}_{t-14}, \widehat{r}_{t-14} \cdot \big(\sum_{s=0}^6 RKI_{t-s-4}\big)^\ast\big/p_{0,t-14}\big),
    \end{equation}
    using the starting values 
    \[ \big(\sum_{s=0}^6 RKI_{t-s-4}\big)^\ast = \sum_{s=0}^6 RKI_{t-s-4},\]
    for $ t=15, 16, 17 $ and $ 18$.
 \item[]\ {\it Step 1:} \   Calculate  for $ t=15, 16, \ldots, n$ the pseudo estimator 
    \[ \widehat{R}^\ast_t = \frac{\big(\sum_{s=0}^6RKI_{t-s}\big)^\ast}{\big(\sum_{s=0}^6 RKI_{t-s-4}\big)^\ast}. \]
    \item[] {\it Step 3:} \ Repeat Step 1  and Step 2 a large number of  times, say $B$ times,  and denote by
        \[ \widehat{R}^\ast_{t,1},  \ \widehat{R}^\ast_{t,2}, \ldots, \  \widehat{R}^\ast_{t,B} \]
        the pseudo-random variables obtained for $ t \in \{15, 16, \ldots, n\}$.
    \item[] {\it Step 4:} \ For a desired $ 1-\alpha$ confidence level, let
    $ Q_1=[B*\alpha/2]$ and $ Q_2 = [B*(1-\alpha/2)]$. A $(1-\alpha)100\%$ confidence interval for $ R_{0,t}$ is then given by 
    \[  \big[ \widehat{R}^\ast_{t,(Q_1)} , \ \widehat{R}^\ast_{t,(Q_2)}\big], \]
    where $ \widehat{R}^\ast_{t,(1)},  \widehat{R}^\ast_{t,(2)}, \ldots,  \widehat{R}^\ast_{t,(B)} $ 
    denotes the ordered values of the random sample 
    $R_{t,i}^\ast$, $i=1,2, \ldots, B$, generated in Step 3.
\end{enumerate}

Notice that the above algorithm imitates also the dependence in generating  the random sums $ (\sum_{s=0}^6 RKI_{t-s})^\ast$ by using the time dependent parameters $ \widehat{p}_{t-14}$ and $ \widehat{r}_{t-14}$ as  well as the (by four time units) lagged sum  $ (\sum_{s=0}^6 RKI_{t-s-4})^\ast$. Since we need $\widehat{p}_{t-14} $ and $\widehat{r}_{t-14} $, to generate $(\sum_{s=0}^6 RKI_{t-s})^\ast$,  our bootstrap algorithm starts for $ t=15$ and needs  starting values like those given in Step 1. 

A simplified version of the proposed bootstrap algorithm can be applied when interest is focused in the construction of  a confidence interval for a particular time point  $t$ only. In this case $ (\sum_{s=1}^6 RKI_{t-s-4})^\ast$ in equation (\ref{bootS-1}) can be treated as fixed and replaced by  the observed sum $ \sum_{s=0}^6 RKI_{t-s-4}$. Notice that in  this  case, the bootstrap algorithm essentially estimates via Monte Carlo the percentage points of the  Negative Binomial distribution,
$ \mathcal{NB}\big(\widehat{q}_{t-14}, \widehat{r}_{t-14} \cdot \big(\sum_{s=0}^6 RKI_{t-s-4}\big) \big/p_{0,t-14}\big)$. Furthermore, also in this case  and by  construction, 
\begin{align*}
 E^\ast(R^\ast_t)& =  \frac{1}{p_{0,t-14}} \cdot \frac{\widehat{r}_{t-14}\big(1-\widehat{q}_{t-14}\big)}
 {\widehat q_{t-14}} =   \frac{\widehat{r}_{t-14}\big(1-\widehat{p}_{t-14}\big)}
 {\widehat{p}_{t-14}}
 \end{align*}
 see also (\ref{eq-13}), which  is the expected value $ R_{0,t-14}$ of the individual reproduction number $ R_{t-14}$ and the parameter which the estimator  $\widehat{R}_{0,t}^7$ actually  estimates. Thus like the estimator $ \widehat{R}_{0,t}$, the bootstrap   delivers a confidence interval for $R_{0,t-14}$. 

\begin{figure}[htbp]
\begin{center}
\includegraphics[scale=0.5]{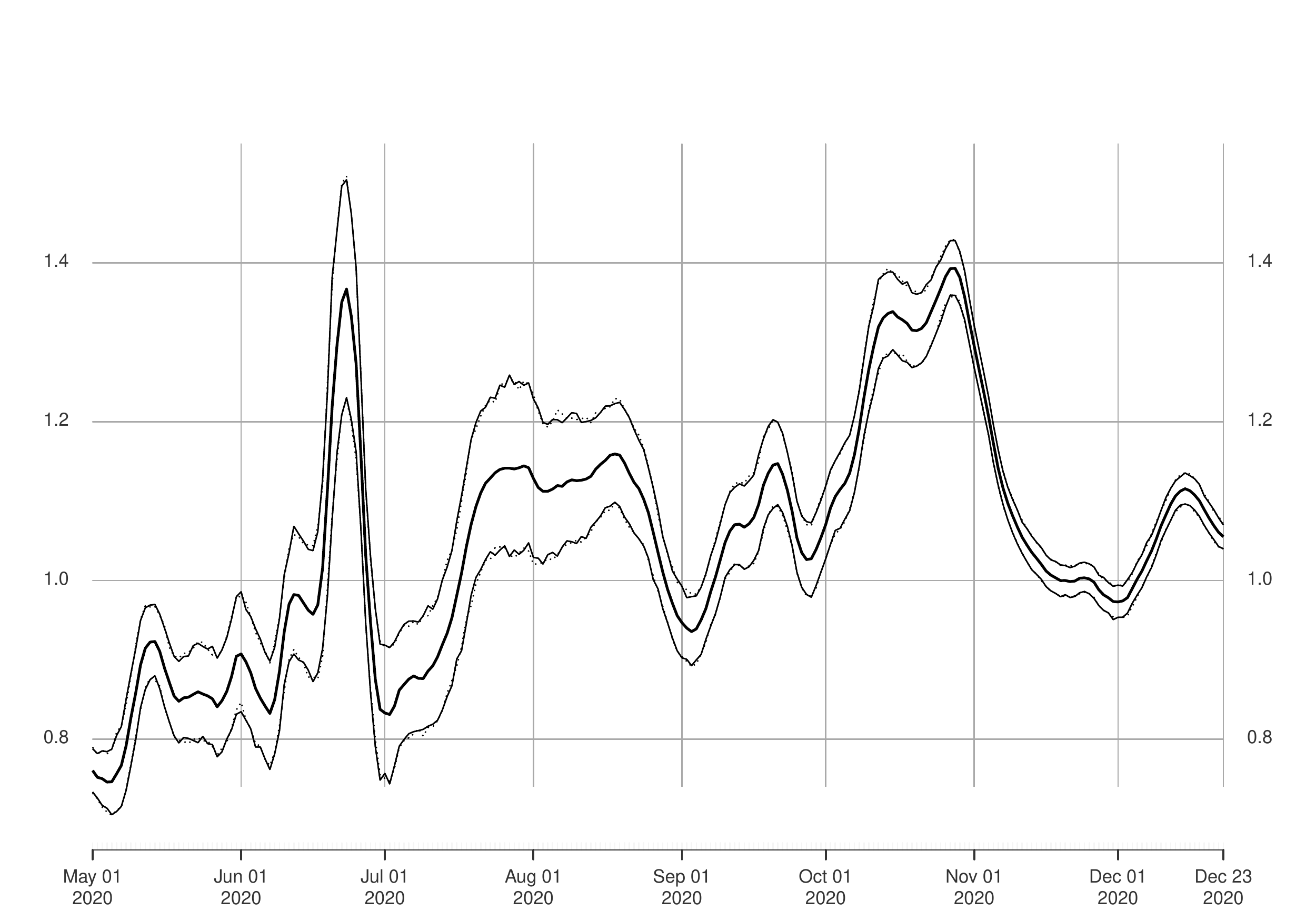}
\caption{Estimates $\widehat{R}_{0,t}^7$ and the $95\%$ pointwise confidence intervals for different reporting rates $p_0=0.2$ (solid) and $p_0=0.5$ (dotted).}
\label{fig:bootstrap}
\end{center}
\end{figure}

Figure \ref{fig:bootstrap} shows estimators $\widehat{R}_{0,t}^7$ obtained from  \COVID-cases reported to RKI in Germany together with the corresponding $95\%$ pointwise confidence intervals constructed using the bootstrap algorithm  proposed in this section.

\section{Applications}
\label{results}

\subsection{Validating the Negative Binomial Hypothesis}
We first investigate the suitability of the assumed Negative Binomial distribution for describing the random behavior of the individual reproduction number using   the number of infections reported by RKI.  Toward this end and as for estimating the variance,  we focus on reported {\COVID}-cases  on district level, i.e., on the observations $ RKI_{t,\ell}$. This  allows for getting replicates of the  random variable of interest  and, therefore,  for  testing the  hypothesis that the individual reproduction number follows a Negative Binomial distribution.  Recall that in our  discussion in Section 3, it was   assumed  that the sum of reported cases over the  time points $t, t-1, \ldots, t-6$  in district $ \ell$, that is, $\sum_{s=0}^6 RKI_{t-s,\ell}$,   satisfies $(\tau=7),$
\[
\sum_{s=0}^6 RKI_{t-s,\ell} \sim \mathcal{NB}(q_{t-14}, r_{t-14}\cdot \sum_{s=0}^6 N_{t-11-s,\ell}). 
\]
Assuming that $p_{0,t}$ remains almost constant in the range of $\tau$ days we get  using (\ref{eq-04aa}) and (\ref{eq-07}) that  \[ \sum_{s=0}^6 N_{t-11-s,\ell}\approx 
p_{0,t-14}\sum_{s=0}^6 N^\prime_{t-11-s,\ell} \approx  p_{0,t-14}\sum_{s=0}^6 RKI_{t-4-s,\ell}.\]
That is, in terms of the observed RKI data, 
the assumption  we have to test translates to 
\begin{equation} \label{eq.test1}
\sum_{s=0}^6 RKI_{t-s,\ell} \sim \mathcal{NB}\big(q_{t-14},\ \frac{1}{p_{0,t-14}}\cdot r_{t-14}\cdot \sum_{s=0}^6 RKI_{t-4-s,\ell}\big). 
\end{equation}
In order to select from the existing data  appropriate samples
for testing   the above assumption, we proceed as follows. We first select all districts for which the average of reported cases at  time points $ t-4, t-5, \ldots, t-11$ is approximately the same. Practically, this means that we consider  
districts  for which 
\begin{equation} 
\label{eq.condTest}
\frac{1}{7} \sum_{s=0}^6 RKI_{t-4-s,\ell} \in [15,25].
\end{equation}
 We have experienced that chosen a number of average daily infections at district level  outside the above interval,  leads to the selection of a relatively small number of districts, that is to a small sample size. Let $ L_{t,S}$ be the total number of districts at time point $t$ satisfying condition \eqref{eq.condTest} and let    $ \{1,2, \ldots, L_{t,S}\}$ be  the corresponding set of districts. From  the  total number of  time points $t$ available, we further only  consider  those  time points, for which $ L_{t,S} \geq 75$. This  ensures that a sufficiently large number  of districts is  available for   testing the hypothesis of interest.  After applying  this selection procedure to the $RKI_{t,\ell}$  data, we end up with a total of $ T=45$ data points $t$   for which the corresponding condition \eqref{eq.condTest} and $ L_{t,S} \geq 75$ is satisfied.

The problem of testing the goodness-of-fit of a Negative Binomial distribution $ \mathcal{NB}(p,r)$ when both parameters  $p $ and $ r$ are unknown, has been considered  by some authors in the literature; see \citeasnoun{Meintanis2005} and \citeasnoun{Bestetal2009}. \citeasnoun{Meintanis2005} proposed a test based on the comparison of the empirical probability generating function with that of the Negative Binomial distribution with estimated parameters. \citeasnoun{Bestetal2009} considered tests based on the  
comparison of third and fourth order moments. In the following  we focus on the test proposed by \citeasnoun{Meintanis2005} but  we also report results for the test proposed by \citeasnoun{Bestetal2009}.

The test statistic proposed by \citeasnoun{Meintanis2005} with suggested parameter $ a=5$,  applied to the selected RKI data, 
$ Y_{t,\ell}=\sum_{s=0}^6 RKI_{t-s,\ell}, \ell \in \{1,2, \ldots, L_{t,S}\}$,
is given by 
\begin{align*}
T_{t,n} & = \frac{1}{L_{t,S}}\Big[ \overline{Y}^2_{L_{t,S}}\sum_{j,k=1}^{L_{t,S}}I(Y^+_{j,k} +5)  -2 \overline{Y}_{L_{t,S}}\sum_{j,k=1}^{L_{t,S}} Y_{t,j}\Big\{ (1+\widehat{\rho}_{L_{t,S}} ) I(Y^+_{j,k} +4) 
- \widehat{\rho}_{L_{t,S}}I(Y^+_{j,k} +5)\Big\} \\
& \ \ \ \ + \sum_{j,k=1}^{L_{t,S}} Y_{t,j}\cdot Y_{t,k}\Big\{(1+\widehat{\rho}_{L_{t,S}})^2I(Y^+_{j,k} +3)+\widehat{\rho}_{L_{t,S}}I(Y^+_{j,k} +5) \\
& \ \ \ \ -2\widehat{\rho}_{L_{t,S}}(1+\widehat{\rho}_{L_{t,S}}) I(Y^+_{j,k} +4)\Big\}\Big],
\end{align*}
where $Y^+_{j,k}= Y_{t,j} + Y_{t,k}$, 
$ I(\beta) = (1+\beta)^{-1}$ 
for $ \beta >-1$ and 
\[ \widehat{\rho}_{L_{t,S}} = \Big(S^2_{L_{t,S}} - \overline{Y}_{L_{t,S}} \Big)\Big/ \overline{Y}_{L{t,S}},
\]
with 
$\overline{Y}_{L_{t,S}}=L_{t,s}^{-1}\sum_{j=1}^{L_{t,S}} Y_{t,j} $
and 
\[S^2_{L_{t,S}}=L_{t,s}^{-1}\sum_{j=1}^{L_{t,S}} (Y_{t,j} - \overline{Y}_{L_{t,S}})^2. 
\]
To obtain critical values of the $T_{t,n}$ test, a parametric bootstrap procedure is used. More specifically, i.i.d. random samples of length 
$ L_{t,S}$ are generated from a
$\mathcal{NB}(\widehat{q}_t,\widehat{r}_t)$ distribution,  where  
\[
\widehat{q}_t=\frac{\widehat{r}_t}{\widehat{r}_t+ \overline{Y}_{L_{t,S}}} \ \ \mbox{and} \ \  \widehat{r}_t = \frac{L_{t,S}^{-1}\sum_{j=1}^{L_{t,S}} Y_{t,j}^2}{S^2_{L_{t,S}} - \overline{Y}_{L_{t,S}}}. 
\]
The distribution of the test statistic $ T_{t,n}$ under the null is then estimated using the distribution of the same test statistic calculated   using  the bootstrap pseudo random sample.

Applying the above test to the  RKI data sets   selected according to the described procedure,  the null hypothesis of a Negative Binomial distribution has been rejected at the $5\%$ level in  only  $8$ out of the $45$ different data sets considered. 
Notice   that qualitatively the same result is obtained, if one uses the test proposed by \citeasnoun{Bestetal2009}. As already mentioned, this test compares   the empirical third and fourth moments with those of the Negative Binomial distribution fitted to the data; see the test denoted by $R^2/Var(R)$ in page 6 of the cited paper. Applying this test leads to a  rejection of the  the null hypothesis at the  $5\%$ level,  in only $2$  out of the $45$ data sets  selected. Our  testing procedures find, therefore, no  evidence against the assumption 
 that the random behavior of the individual reproduction number is governed by a Negative Binomial distribution.    

\subsection{Empirical Results}
\begin{figure}[bhtp]
\begin{center}
 \includegraphics[scale=0.5]{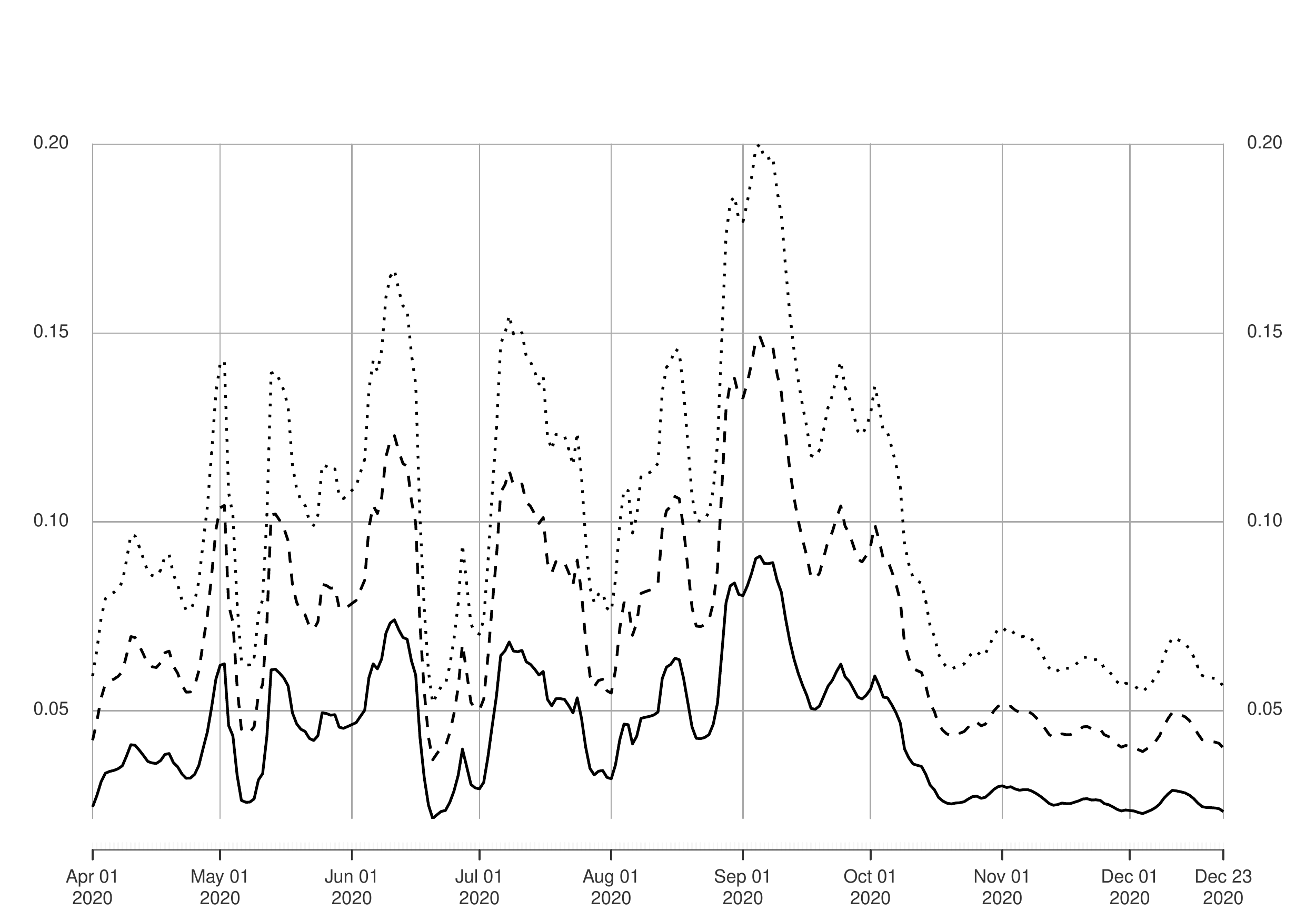}
\caption{The estimated parameter $p_t$ for $p_0=0.2$ (black solid line), $p_0=0.35$ (black dashed line), $p_0=0.5$ (black dotted line).}
\label{plot-pt}
\end{center}
\end{figure}

\begin{figure}[htbp]
\begin{center}
 \includegraphics[scale=0.5]{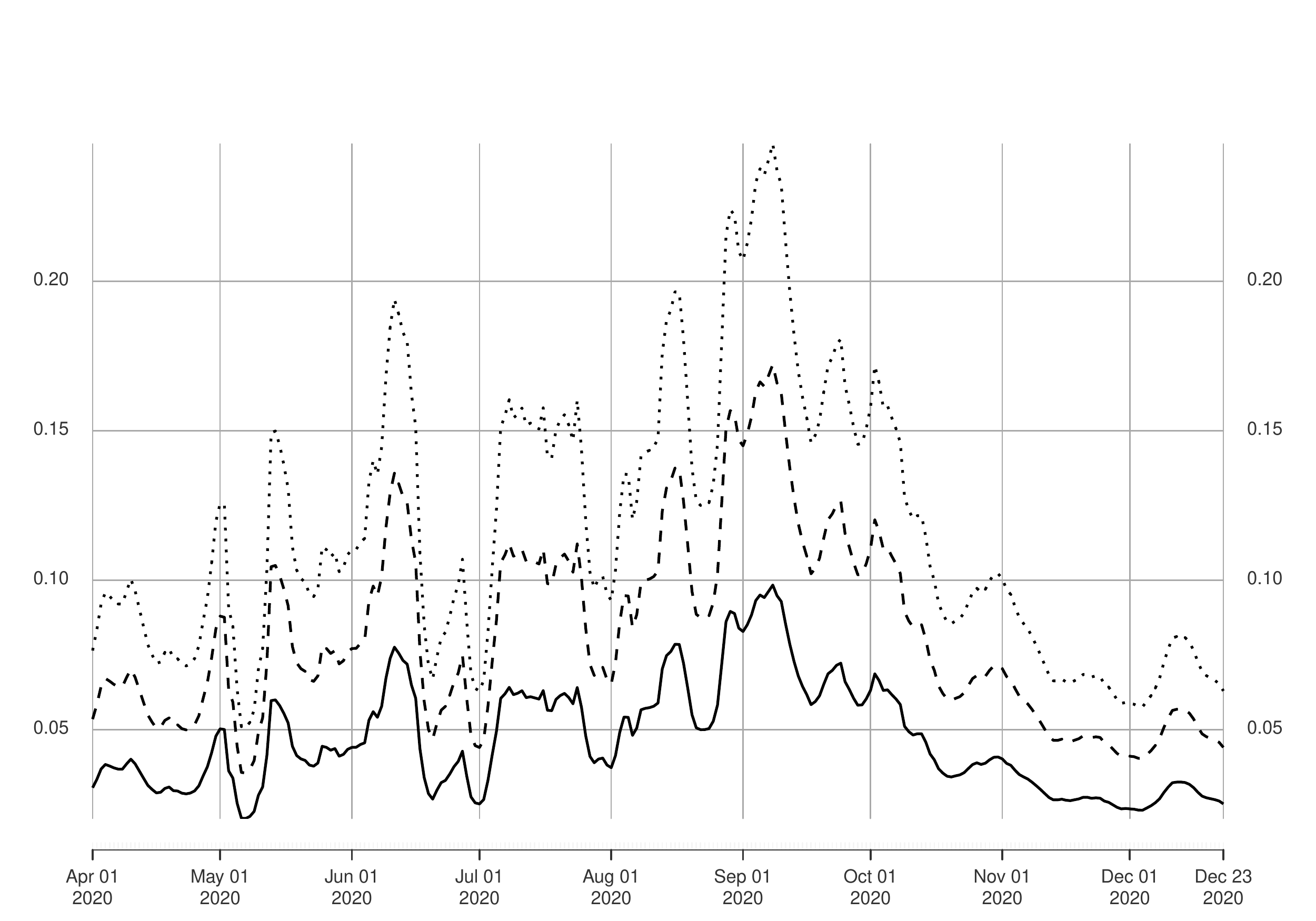}
\caption{The estimated parameter $r_t$ for $p_0=0.2$ (black solid line), $p_0=0.35$ (black dashed line), $p_0=0.5$ (black dotted line).}
\label{plot-rt}
\end{center}
\end{figure}

\begin{figure}[htbp]
\begin{center}
 \includegraphics[scale=0.5]{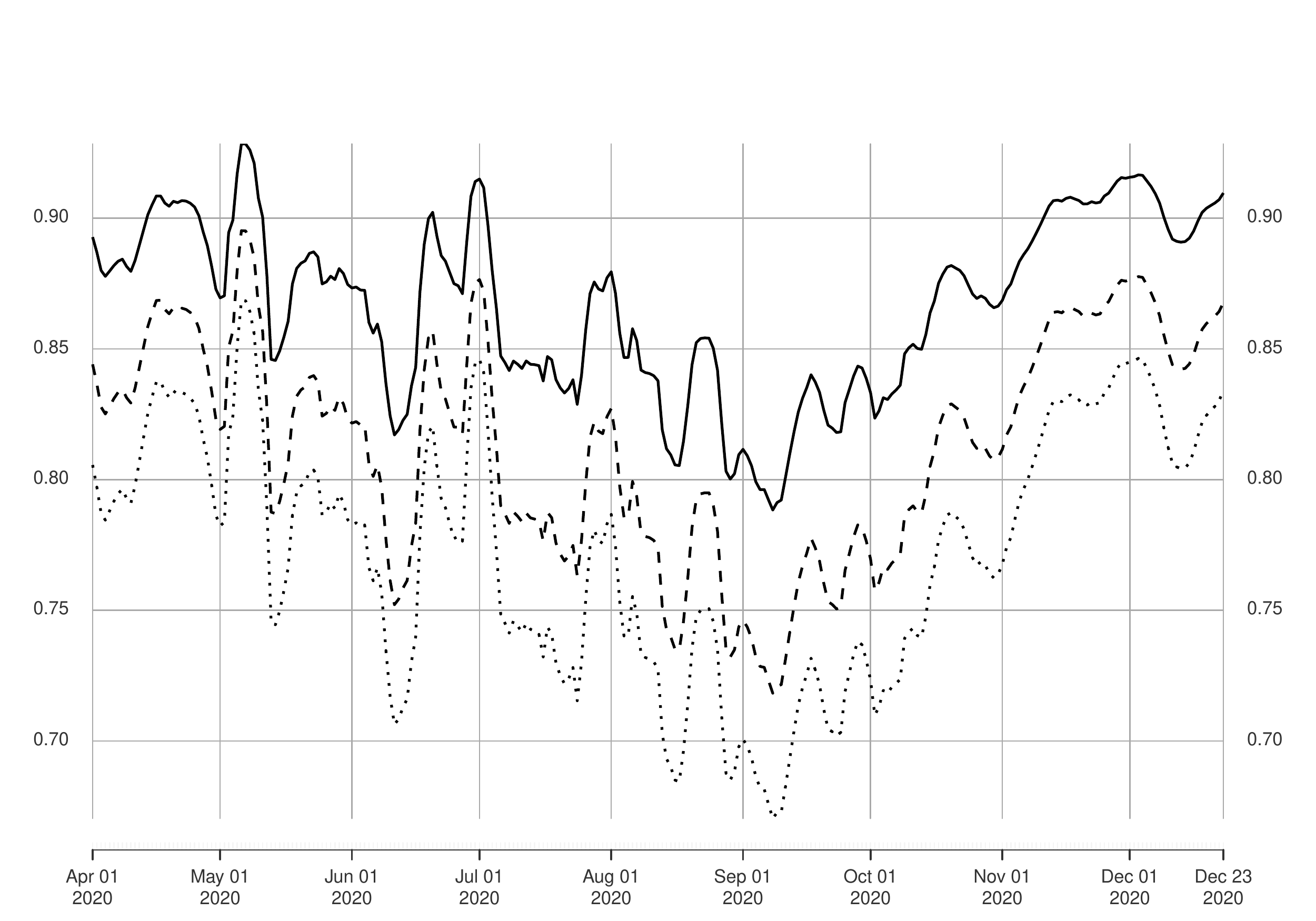}
\caption{The probability that an individual causes no secondary infection for $p_0=0.2$ (black solid line), $p_0=0.35$ (black dashed line), $p_0=0.5$ (black dotted line)..}
\label{plot-P-0}
\end{center}
\end{figure}

\begin{figure}[htbp]
\begin{center}
 \includegraphics[scale=0.5]{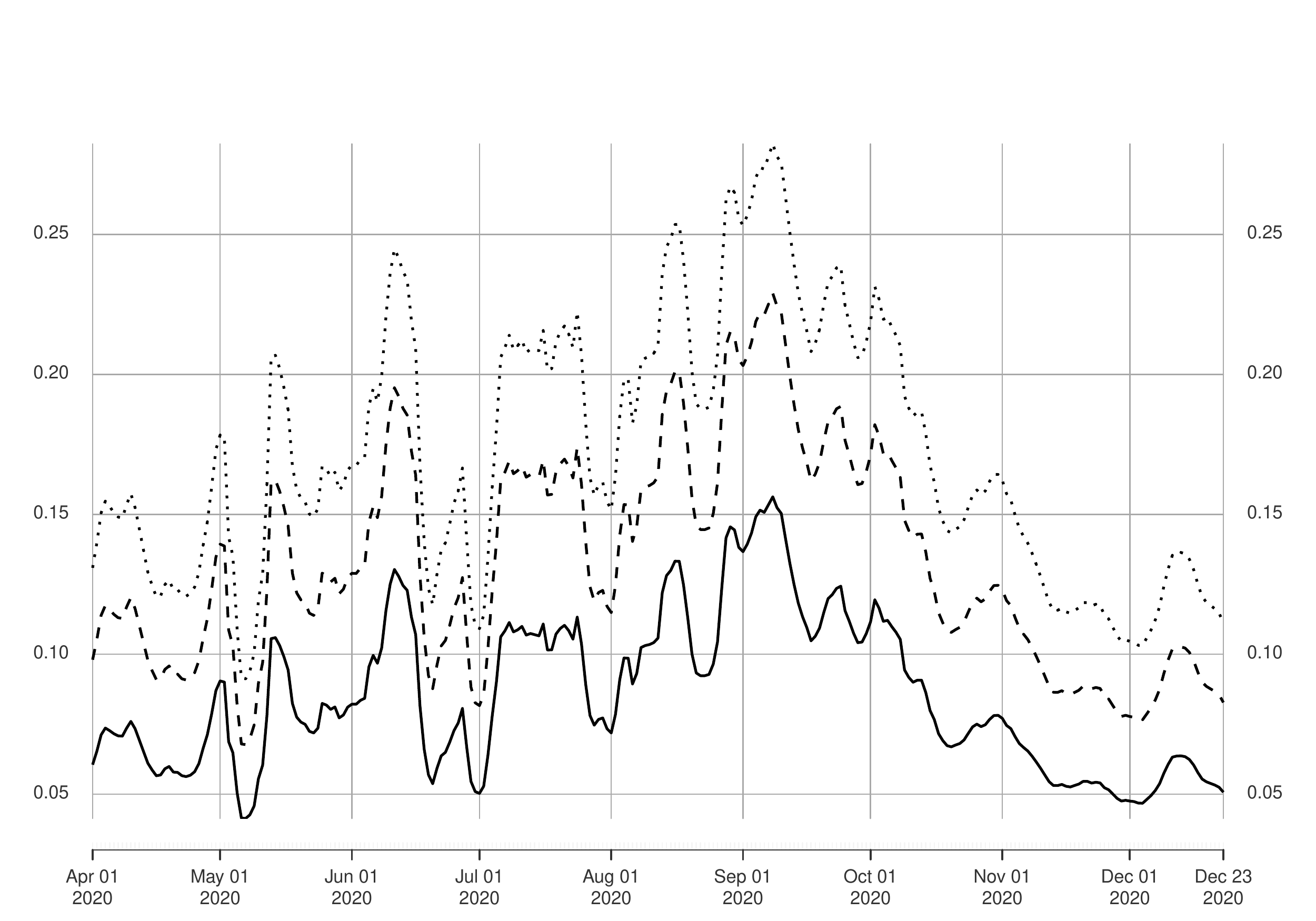}
\caption{The probability that an individual causes between 1 and 5 secondary infections for $p_0=0.2$ (black solid line), $p_0=0.35$ (black dashed line), $p_0=0.5$ (black dotted line)..}
\label{plot-P-1-5}
\end{center}
\end{figure}

\begin{figure}[htbp]
\begin{center}
 \includegraphics[scale=0.5]{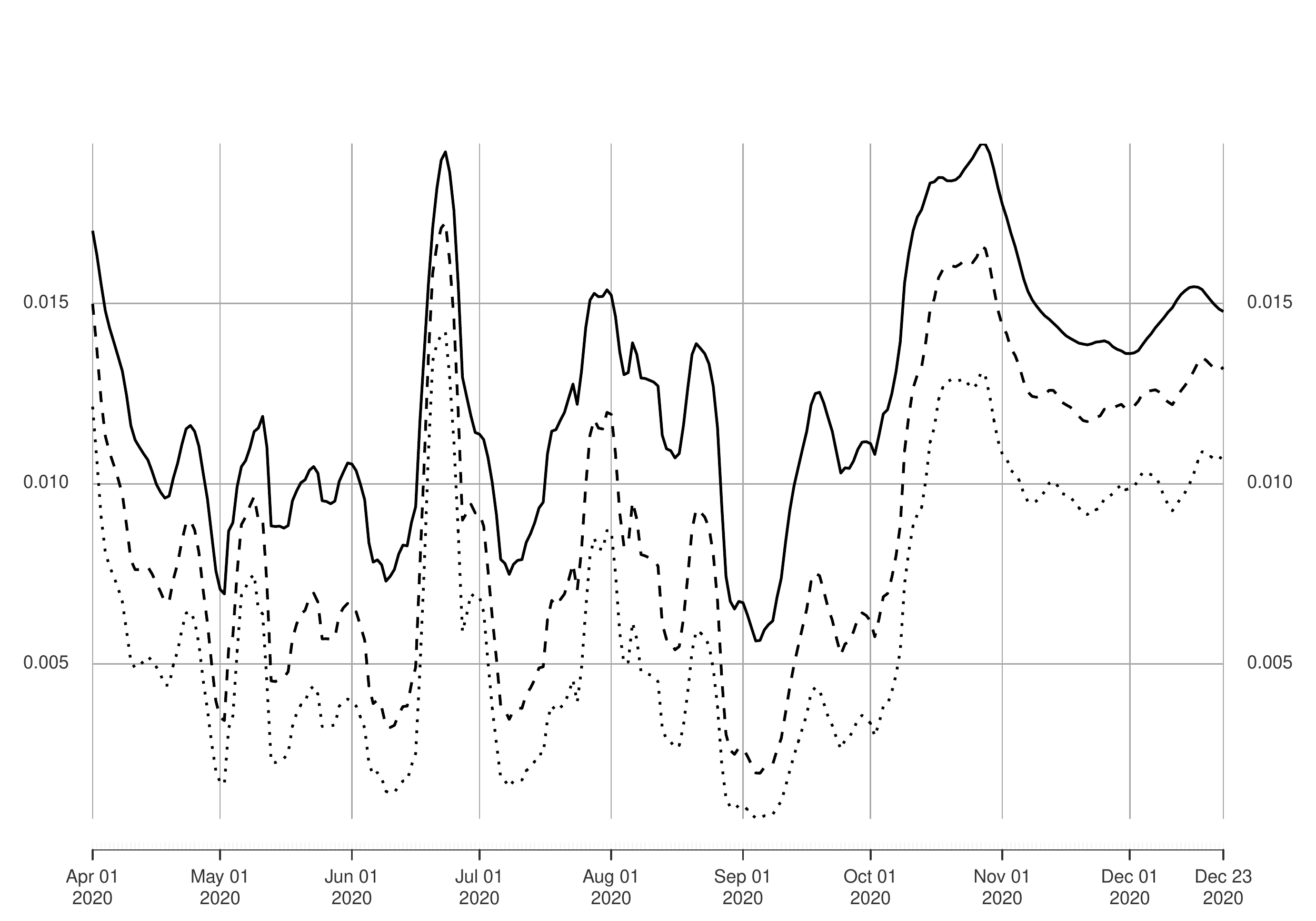}
\caption{The probability that an individual causes 20 or more secondary infections for $p_0=0.2$ (black solid line), $p_0=0.35$ (black dashed line), $p_0=0.5$ (black dotted line)..}
\label{plot-P-20}
\end{center}
\end{figure}

  In this section we present the estimated parameter of the Negative Binomial distribution obtained for Germany based on the RKI data set by the method described in Section~\ref{Estimation}. We consider here the time period from April 1, 2020 to December 23, 2020. Furthermore, the moment estimators in Section~\ref{Estimation} are based on 401 districts and a left-sided 7-day moving average smoothing is applied to the estimated moments before computing the parameter estimators \eqref{eq-18c}. As mentioned in Section~\ref{Estimation}, the parameter estimates depend on the unknown reporting rate $p_{0,t}$ and this  rate cannot be estimated from  the  data given. That is why we present results here for three possible reporting rates, i.e.,  $p_{0,t} \in \{0.2, 0.35, 0.5\}$. For a given reporting rate, the estimated parameters are presented here as connected lines. However, note that the reporting rate may vary over time and the proposed estimation procedure requires only a locally constant reporting rate, i.e., the estimation at time $t$ requires  roughly no substantial changes in the reporting rate for the past three weeks. This means, it is possible to switch over time between the results of different reporting rates, if there is strong believe that the reporting rate changes between different time periods. The estimated parameters $p_t$ and $r_t$ are given in Figure~\ref{plot-pt} and ~\ref{plot-rt}, respectively. In this time period, $p_t$ is in the range of $0.021$ to $0.2$ and $r_t$  in range of $0.02$ to $0.25$.  The parameters also can be translated  into probabilities that an individual case causes a given number of secondary infections over its entire infectious period. For this, we present in Figure~\ref{plot-P-0} the probability that an individual causes no infections, in Figure~\ref{plot-P-1-5} the probability that an individual causes one to five infections, and in Figure~\ref{plot-P-20} the probability that an individual causes 20 or more infections. 
  
  Before discussing the results, first note that over the entire period non-pharmaceutical measures were in place such as mandatory wearing of
  face masks in public areas, detected cases and contacts were quarantined, etc. This also can be seen  in the estimates of the parameter  $R_0$. Over this time period, the average is $1.036$, and consequently, far less than the  reproduction rate without any measures, which is estimated as $3.32$ by \citeasnoun{Alimohamadi2020} in a meta-study.
  Over the entire time period a strong overdispersion can be observed irrespective of the reporting rate. Smaller reporting rates lead in general to smaller parameter values for $r_t$.   Furthermore, in the summer period larger parameter values for $r_t$ can be observed than in the fall period. The overdispersion can be displayed well in in probabilities. During the summer period the probability that an individual cases causes no infection is given by  $60\%-80\%$ and it rises in the fall period to $80\%-90\%$. Additionally, the probability that an individual case causes 20 or more infections almost doubles from summer to fall and peaks in October with values about $1.5\%-2\%$. In contrast, the probability that an individual case causes one to five infections almost halves from summer to fall with values in summer of about $10\%-20\%$.
  
  \citeasnoun{Endo2020} estimated the overdispersion parameter $r_t$ of the Negative Binomial distribution as $0.1$ with a $95\%$ confidence interval from $0.04$ to $0.2$. Note however, that their considered time period is January and February of 2020. In that time period non-pharmaceutical measures such as obligatory face masks in public areas were not yet in place or less strict than in the time period considered here. Since in our considered time period non-pharmaceutical measures were less strict in Germany during the summer 2020, it seems most reasonable to compare the results obtained in the summer period, June 1, 2020 to October 31, 2020, with the results obtain by \citeasnoun{Endo2020}. In the summer period, $r_t$ takes values in the range of $0.025$ to $0.22$ and for a reporting rate of $0.35$, we obtain values in the range of $0.044$ to $0.157$ with a mean value of $0.095$. Hence, the obtained values coincide well with the values obtained in  \citeasnoun{Endo2020}.
  
\FloatBarrier

\subsection{Simulated Interventions - Cluster Tracing} \label{sec:simulations}
The estimated parameters can be used to simulate the effect of additional non-pharmaceutical measures such as additional cluster tracing or physical distancing. This means  that the number of daily infections is simulated using the obtained parameter estimates and additional interventions can be plugged-in. Note that non-pharmaceutical measures already in place during the time period used for estimation are reflected in the estimated parameters.
As an example, we consider cluster tracing. Let $S_{i,t}$ denotes the number of secondary infections caused by an individual case $i$ at time $t$ and $\widetilde S_{i,t}$ the number of cases caused by this person and observed by the health authorities. The health authorities are tracing a cluster if $\widetilde S_{i,t}>CS$, where $CS$ denotes the considered cluster size. They are able to prevent the traced cases from causing further infections by isolating them with some effectiveness $C_{eff}\in[0,1]$. This means, if $\widetilde S_{i,t}>CS$ only $S_{i,t}-C_{eff}\widetilde S_{i,t}$ secondary cases can cause further infections. 

\begin{figure}[htbp]
\begin{center}
 \includegraphics[scale=0.5]{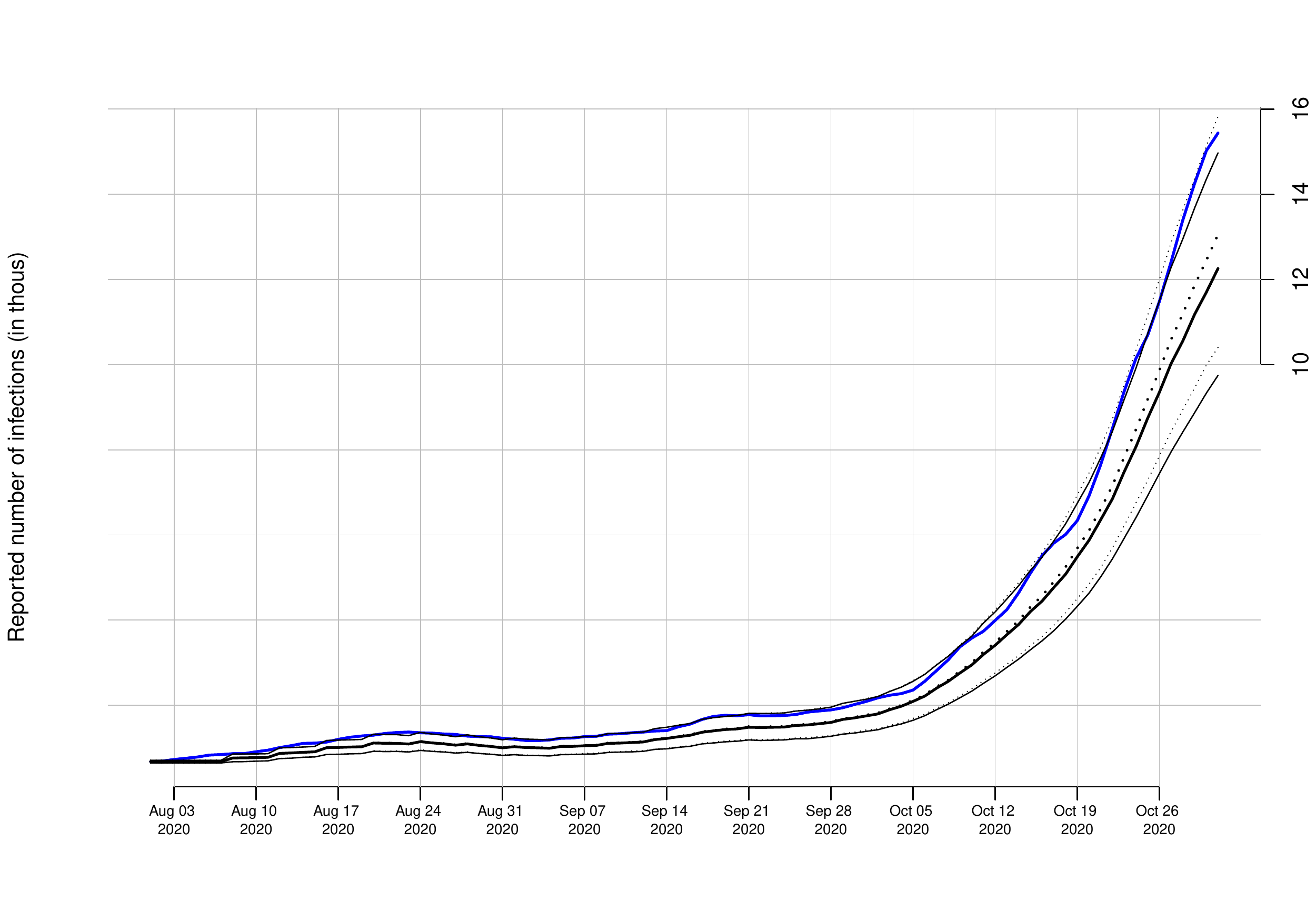}
\caption{The effect of cluster tracing on the reported number of daily infected persons. Observed cluster with size $20$ or greater are traced with effectiveness of $C_{eff}=0.35$. The base case is given in solid blue. Different reporting rates ($p_0=0.2, 0.5)$ are given in black solid and dotted lines respectively. The simulation is based on 10,000 trials and the mean case is given by thick lines whereas the upper and lower $5\%$ cases are given by thin lines. 
}
\label{Plot-Szen-1}
\end{center}
\end{figure}

\begin{figure}[htbp]
\begin{center}
 \includegraphics[scale=0.5]{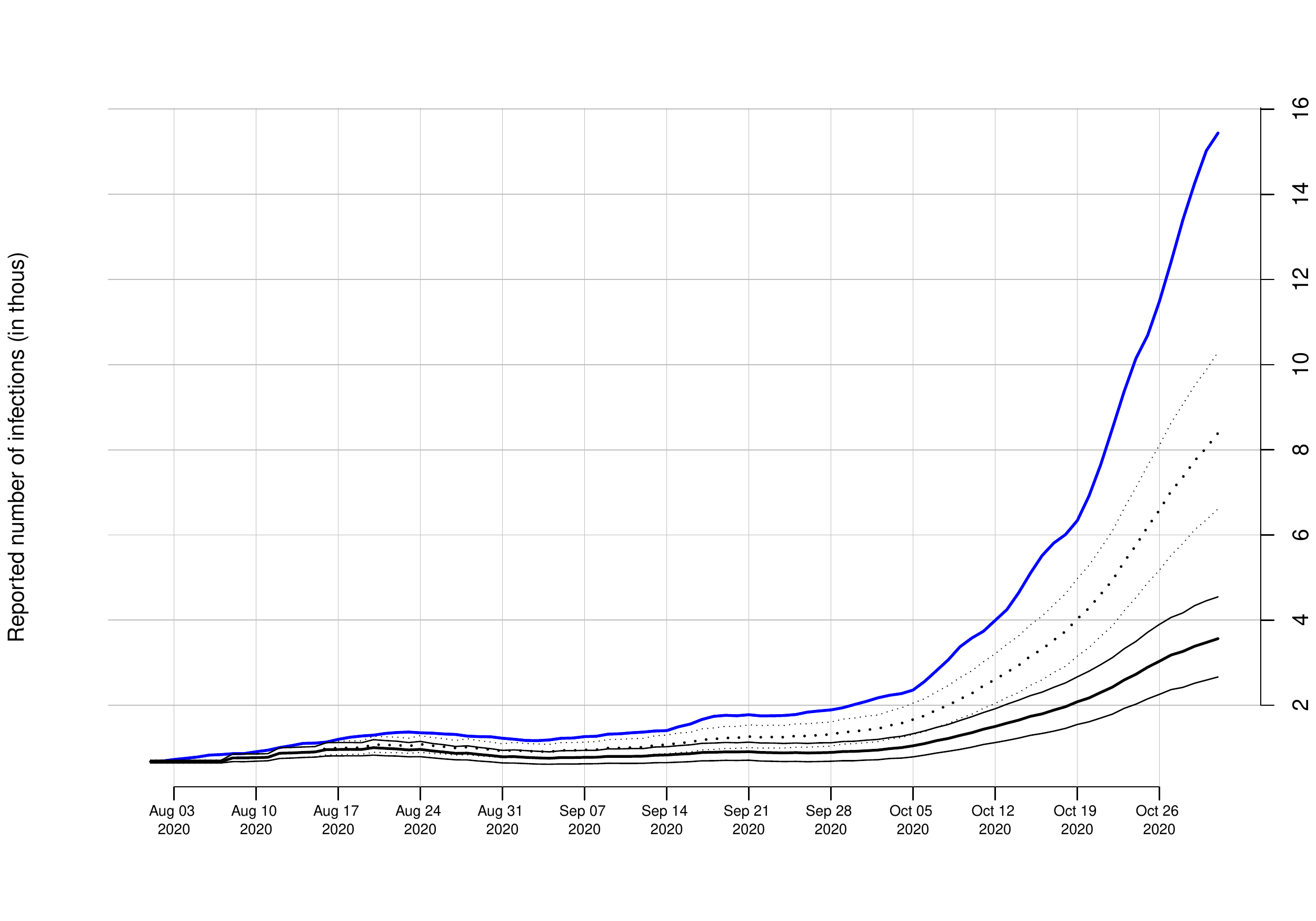}
\caption{The effect of cluster tracing on the reported number of daily infected persons. Observed cluster with size $5$ or greater are traced with effectiveness of $C_{eff}=0.35$. The base case is given in solid blue. Different reporting rates ($p_0=0.2, 0.5)$ are given in black solid and dotted lines respectively. The simulation is based on 10,000 trials and the mean case is given by thick lines whereas the upper and lower $5\%$ cases are given by thin  lines.}
\label{Plot-Szen-2}
\end{center}
\end{figure}

We consider here the time period August 1, 2020 to October 31, 2020. With the beginning of November stricter non-pharmaceutical measures have  been active in Germany. Since identified cases are usually persons with symptoms of {\COVID} and a positive laboratory test of {\SARS}, there is possible a time delays  before the health authorities can isolate the persons of a cluster. That is why, we set the effectiveness of cluster tracing to $C_{eff}=0.35$. For the cluster size we consider two cases: observed clusters with size $20$ or greater are traced and observed clusters with size $5$ or greater are traced. These two cases are displayed in Figure~\ref{Plot-Szen-1} and \ref{Plot-Szen-2}, respectively. We consider two possible reporting rates $p_{0,t} \in [0.2,0.5]$. Furthermore, the simulation is based on 10,000 trials and the mean case is given by thick lines and the upper and lower $5\%$ cases by thin lines. 

For the case that observed cluster with size $20$ or greater are traced with effectiveness of $C_{eff}=0.35$, the rise in daily infections during September and October can only be delayed by one week. For this case, the reporting rate does not affect much the results. This changes for the case where  observed clusters with size $5$ or greater are traced with effectiveness of $C_{eff}=0.35$. In this case, the rise in daily infections during September and October can only by delayed by two weeks for a reporting rate of $0.2$ and for reporting rate of $0.5$ the corresponding delay is almost one month. However, even in the most optimistic case, i.e., a reporting rate of $50\%$ and clusters with size $5$ or greater are being traced, a strong increase in daily infections during September and October cannot be stopped.

\FloatBarrier
 \section{Appendix}
 \begin{lem} Let $ X\sim \mathcal{NB}(p,r)$ with parameters $ p\in(0,1)$ and $ r>0$.  If $ Z_1, Z_2, \ldots $ are i.i.d. $Bernoulli(p_0)$ variables, then
 \[ Y:= \sum_{j=1}^XZ_j \sim \mathcal{NB}(q,r) \ \ \mbox{with} \ \  q=\frac{p}{p+p_0-p_0\cdot p}.\]
 \end{lem}
 {\it Proof:} \  Notice first that for $X=n$ given, the distribution of $ Y|X=n$ is $ Binomial(n,p_0)$. Furthermore, if  $ X\sim \mathcal{NB}(p,r)$ then $ X \sim Poisson(\lambda) $ with $ \lambda \sim Gamma\big(r, p/(1-p)\big)$; see \eqref{eq-01}. From these we get for $ k\in \N\cup\{0\} $,
 \begin{align*}
P(Y=k) & = \sum_{n=k}^\infty P(Y=k| X=n)\cdot P(X=n)\\
& =\sum_{n=k}^\infty \left(\begin{array}{c} n\\ k \end{array}\right)p_o^k(1-p_0)^{n-k} \int_0^\infty \frac{e^{-\lambda} \lambda^n}{n!} \cdot f_{r,\frac{p}{1-p}}(\lambda)d\lambda\\
& =\int_0^\infty \frac{(p_0\cdot\lambda)^k e^{-\lambda}}{k!}
\sum_{s=0}^\infty\frac{(1-p_0)^s\lambda^s}{s!}\cdot f_{r,\frac{p}{1-p}}(\lambda)d\lambda\\
& = \int_0^\infty\frac{(p_0\cdot\lambda)^k e^{-p_o\cdot \lambda}}{k!}
\cdot f_{r,\frac{p}{1-p}}(\lambda)d\lambda\\
& = \int_0^\infty\frac{\widetilde\lambda^k e^{-\widetilde\lambda}}{k!}
\cdot \frac{1}{p_0}f_{r,\frac{p}{1-p}}
\Big(\frac{\widetilde\lambda}{p_0}\Big)d\widetilde\lambda,
 \end{align*}
where the last equality follows using the substitution $ \widetilde{\lambda}=p_o\cdot\lambda$. Since 
\begin{align*} 
\frac{1}{p_0}f_{r,\frac{p}{1-p}}
\Big(\frac{\widetilde\lambda}{p_0}\Big) 
& = \frac{1}{\Gamma (r)}\Big(\frac{p}{(1-p)p_0}\Big)^r \widetilde{\lambda}^{r-1}
e^{-\widetilde{\lambda} \frac{p}{(1-p)p_0} } \\
& = f_{r,\frac{q}{1-q}}
\big(\widetilde\lambda\big),
\end{align*}
 where $ q = p/(p+p_0-p_0\cdot p)$,
we  get  
\[ P(Y=k) = \int_0^\infty\frac{\widetilde\lambda^k e^{-\widetilde\lambda}}{k!}
\cdot f_{r,\frac{q}{1-q}}
(\widetilde\lambda)d\widetilde\lambda,\]
which is the probability function of the $ \mathcal{NB}(q,r)$ distribution.
\flushright{$\Box$}
 
  

\end{document}